# Objective Reduction of the Wave Function Demonstrated on Superconducting Quantum Computer

James Tagg, William Reid
Valis Corporation, Encinitas, California, 92024
james.tagg@valiscorp.com

We describe an experiment using superconducting transmon qubits that demonstrates wavefunction collapse consistent with Orchestrated Objective Reduction (Orch-OR)—the theory of consciousness proposed by Sir Roger Penrose and Stuart Hameroff. The experiment performs a partial measurement on a qubit. It uses the result of that measurement to move an estimated $10^{-12}$ kg of mass in one of two locations separated by approximately 1 mm. In standard quantum mechanics, the partial measurement leaves the system in an *improper mixture*—a state that appears probabilistic but retains quantum coherence. This is mathematically indistinguishable from a *proper mixture*, where the state has genuinely collapsed. According to Penrose's theory, improper mixtures can lead to gravitationally induced collapse. Results of our experiment show a change in the circuit evolution that is consistent with wavefunction collapse driven by such an improper mixture. The experiment is performed on an IBM Eagle 127-qubit processor using IBM's programming framework Qiskit.

*Introduction*

Schrödinger's famous thought experiment involves a cat in a sealed box with a radioactive sample and poison. Radioactive decay—a quantum event—triggers the release of the poison, killing the cat. Quantum mechanics says the cat is simultaneously alive and dead. Schrödinger thought this was absurd and meant quantum mechanics was incomplete. Einstein agreed.

Lajos Diósi [1–4] and Roger Penrose [5–7] propose a solution where gravity causes the wavefunction to collapse to avoid the paradox of two incompatible spacetimes. In Schrödinger's experiment, one might imagine the live cat standing while the dead cat falls to the floor, giving space-time two different curvatures. The paradox ends when the incompatibility becomes certain and exceeds the time-energy version of the Heisenberg uncertainty principle. Penrose and Diósi derive the equation for the collapse in two ways, differing by a factor of two in the self-energy gamma (γ) constant; hence, their interpretation is called the Diósi-Penrose model [8].

*Penrose Derivation via General Covariance*

Penrose argues that the superposition principle of quantum mechanics breaks down for systems involving significantly different spacetime geometries.

Invariance dictates that physical laws remain unchanged under coordinate transformation. The standard Schrödinger equation in quantum mechanics for a free particle is:

$$i\hbar \frac{\partial}{\partial t}\psi(x,t) = -\frac{\hbar^2}{2m}\nabla^2\psi(x,t) \qquad (1)$$

Coordinates transform as:

$$x' = x - vt, t' = t \qquad (2)$$

The wavefunction transforms accordingly:

$$\psi'(x',t') = \exp\left[\frac{im}{\hbar}\left(v \cdot x - \frac{1}{2}v^2 t\right)\right]\psi(x,t) \quad (3)$$

Penrose considers a superposition of two spatially separated mass distributions, each generating a distinct gravitational field and corresponding spacetime geometry. The gravitational potential energy between these two distributions is given by:

$$E_G = -G\int \frac{\rho(x)\rho(y)}{|x-y|} d^3x d^3y \quad (4)$$

where G is the gravitational constant, and $\rho$(x) is the mass density.

For two specific mass configurations |α⟩ and |β⟩, the gravitational self-energies are:

$$E_{G,\alpha} = -G\int \frac{\rho_\alpha(x)\rho_\alpha(y)}{|x-y|} d^3x\, d^3y \quad (5)$$

$$E_{G,\beta} = -G\int \frac{\rho_\beta(x)\rho_\beta(y)}{|\, x-y \,|} d^3x\, d^3y$$

In the superposition state:

$$|\,\Psi\rangle = a \,|\, \alpha\rangle + b \,|\, \beta\rangle \quad (6)$$

each component evolves independently via:

$$i\hbar\frac{\partial}{\partial t} \,|\, \Psi\rangle = \hat{H} \,|\, \Psi\rangle \quad (7)$$

resulting in:

$$|\,\Psi(t)\rangle = ae^{-iE_\alpha t/\hbar} \,|\, \alpha\rangle + be^{-iE_\beta t/\hbar} \,|\, \beta\rangle \quad (8)$$

This leads to an inconsistency in general relativity since the system exists in a superposition of incompatible spacetime geometries.

To resolve this, Penrose proposes an objective reduction (OR) that occurs after a characteristic time $\tau$, depending on the gravitational self-energy difference:

$$\tau = \gamma\frac{\hbar}{E_g} \quad (9)$$

Where $E_g$ is the gravitational self-energy of the superposition, ħ is the reduced Plank constant, and $\gamma$ is a constant originally estimated by Penrose to be 1/(8π).

This is the same energy released when dust clouds coalesce and 'fall' into the gravitational well to form a planet. If we wish to put our planet into superposition, we must work against this enormous energy. We should immediately say that Penrose is *not* arguing we need to find this energy. That would conflict with conservation laws. Instead, when it becomes certain we *need* to find this energy, the wavefunction collapses instead.

This time dependence elegantly solves Schrödinger's paradox. A proton separated by its own radius would collapse in $10^7$ years, a dust particle in $10^{-8}$ seconds, and a cat in approximately $10^{-28}$ seconds [9]. We never observe superpositions of alive and dead cats because they are far too short-lived to perceive.

We do not know the exact self-energy of the mass involved in a transmon quantum computer, as it is composed of silicon chips, waveguides, and microwave emitters. However, we can estimate the effect by assuming mass elements are separated by ~1 mm and, crucially, do not overlap. Approximately 50% of the gravitational self-energy comes from separating the elements by their radius, and the other 50% from extending the separation to infinity.

IBM transmon qubits lose 50% of their coherence in ~100 µs. To observe Penrose collapse, we need to wait long enough—but not too long—to allow OR effects to develop. This sets a "Goldilocks" delay time of around 50 µs.

Furthermore, cross-talk between qubits in transmon systems means every qubit is partially entangled with others, and if coupling was too great, the system would collapse.

Since mid-circuit partial measurement are supported in IBM's architecture, this gives us confidence that the mass of the measurement system lies within the OR-sensitive range.

| Bits ↓ | $10^{-15}$ kg | $10^{-14}$ kg | $10^{-13}$ kg | $10^{-12}$ kg | $10^{-11}$ kg | $10^{-10}$ kg |
|---|---|---|---|---|---|---|
| 2 bits | (>40 s) | (400 ms) | (4 ms) | (40 μs) | (400 ns) | (4 ns) |
| 4 bits | (>10 s) | (100 ms) | (1 ms) | (10 μs) | (100 ns) | (1 ns) |
| 8 bits | (2.5 s) | (25 ms) | (250 μs) | (2.5 μs) | (25 ns) | (250 ps) |
| 16 bits | (650 ms) | (6.5 ms) | (65 μs) | (650 ns) | (6.5 ns) | (65 ps) |
| 32 bits | (160 ms) | (1.6 ms) | (16 μs) | (160 ns) | (1.6 ns) | (16 ps) |
| 64 bits | (40 ms) | (400 μs) | (4 μs) | (40 ns) | (400 ps) | (4 ps) |

*Fig. 1: Collapse time (τ) explicitly computed for fixed spatial separation $10^{-4}$ m (100 μm). • Green cells: Explicitly optimal experimental regime (~50 μs). • Yellow cells: Feasible but suboptimal (10 μs–500 μs). • Red cells: Clearly outside practical feasibility (<10 μs or >500 μs). The mass is unknown. IBM quantum computers decohere 66% after 150us, and Orch-OR builds up over time; therefore, we search from 10-100us.*

*Principle of the Experiment*

We set up an experiment to perform a partial measurement on a test qubit, which we then treated as an *improper mixture* [10]. The mass involved in measuring a quantum state—whether on an IBM quantum computer or any superconducting quantum platform—is well below the Penrose limit, i.e., below the threshold at which gravitationally induced collapse is expected to occur.

The problem with an improper mixture is that it is mathematically indistinguishable from a *proper mixture*. To illustrate, consider a simple analogy:

If I ask you to flip a coin and immediately tell me the result, you'll get heads or tails with equal 50:50 probability. If instead, you flip the coin, place it in your pocket, and show me the result tomorrow, the statistics are still 50:50. From a probabilistic standpoint, they are identical.

However, from an *engineering* point of view, these two scenarios differ. The coin in your pocket warms to your body's temperature may experience environmental noise, and, crucially, has time to decohere. Irrelevant for a macroscopic coin but relevant if it were a microscopic coin. From Penrose's perspective, this time allows the system to transition from an improper to a proper mixture—due to gravitational collapse.

In our experiment, we take advantage of this potential difference. We use the improper mixture to control the flipping of two qubits. The logic is as follows: if the test qubit's partial measurement result is 1, we flip *gravity_qubit_1*; if it's 0, we flip *gravity_qubit_2*. We call them "gravity qubits" descriptively—not because the qubits themselves have significant mass, but because the classical control electronics (e.g., microwave pulse generators) used to flip them do. The associated electronics form a quantum-gravity bit (gqubit).

If the test qubit is in an improper mixture, this process places the associated control systems into a quantum superposition. These physical components are relatively massive and separated by several millimeters—evident in published images of IBM quantum hardware. If the qubit had instead collapsed to a proper mixture, no such gravitational superposition would occur.

*Experimental Setup*

The experimental protocol uses two primary qubits designated as Control and Test, two ancillary qubits for partial measurement, and two additional qubits that represent gravitational influence (referred to as Gravity Bits). The experiment proceeds in the following stages:

1. Initialization and Superposition Creation: The Control and Test qubits are initialized in the $|0\rangle$ state, flipped into $|1\rangle$ using X gates, and then placed into coherent superposition using Hadamard gates.
2. Partial Measurement via CRY Gates: Controlled-RY gates (CRY) with a rotation angle of $\pi/4$ are used to entangle each primary qubit with its respective ancilla. This implements a weak (partial) measurement. The ancilla qubits are immediately measured to generate what Qiskit labels as classical bits.
3. Conditional Gravity Interaction: Based on the partial measurement of the Test qubit, gravity bits are conditionally flipped, the gravitational self-energy of the microwave subsystems should cause collapse of the system.
4. Quantum Delay and Final Check:

A delay of 50 microseconds is applied, allowing potential gravitational effects to manifest as predicted by Orch-OR theory.

5. Final Hadamard gates applied to both Control and Test qubits translate coherence into measurable probabilities.

*Measurement and Analysis*

Final measurements of the Control and Test qubits are conducted to determine the impact of gravitational bit flips on quantum coherence and collapse probabilities. The gravity bits themselves remain unmeasured to maintain the experiment's coherence dynamics.

*Principle of the Experiment*

Standard quantum mechanics predicts no difference between the Control and Test bits after a partial measurement. Once a qubit's state has been measured and the result treated classically, no further operation on that classical bit should influence the qubits.

Orch-OR, in contrast, predicts that if a quantum system evolves into a superposition of sufficiently massive configurations, the entire system—including the original entangled parts—will undergo objective collapse. The collapse depends on the amount of mass involved and the duration of evolution. If those parameters fall within the right range, collapse should be observable.

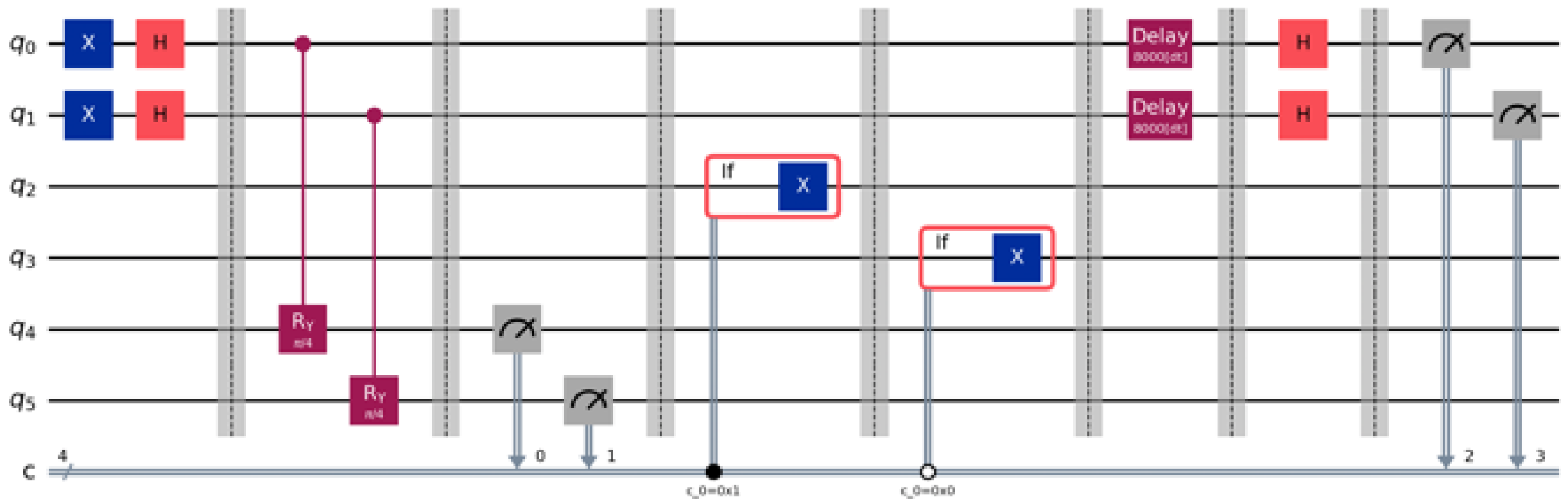

*Figure 2: Quantum circuit for the experiment (two gravity-bit version). The control and test qubits are initialized and placed into superposition using Hadamard gates. Each undergoes a partial (weak) measurement via a CRY gate and measurement onto a classical bit. After a short delay (~50 µs), the classical result from the test qubit is used to conditionally flip one of two gravity qubits. If the classical result is 0, gravity_qubit_1 is flipped; if it's 1, gravity_qubit_2 is flipped. This creates a Schrödinger-style superposition in the associated mass configuration of the classical hardware. Standard Quantum Mechanics (SQM) predicts that operations following a classical measurement cannot affect quantum coherence upstream. Orch-OR, however, predicts that a gravitational superposition will induce collapse of the full system over time. The collapse dynamics depend on the mass-energy difference $E_g$ and coupling time $\tau$. The control qubit serves as an internal reference, never participating in the superposed mass state.*

*The Quantum Circuit*

For the two-bit version of the circuit, the configuration is as follows. Larger versions scale up the number of gravity bits in parallel.

Quantum Bits:

- Qubit 0 **(**Control Qubit**):** Initialized to |0⟩, will be manipulated and measured.
- Qubit 1 **(**Test Qubit**):** Initialized to |0⟩, will be manipulated and measured.

- Qubit 2 **(**Gravity Bit**):** Initialized to $|0\rangle$, flipped conditionally based on partial measurement outcome. This qubit's control electronics have associated physical mass.
- Qubit 3 **(**Gravity Bit**):** Also initialized to $|0\rangle$, with control electronics spatially separated by a few millimeters from Qubit 2, forming a distinct mass configuration.

These gravity bits refer not to the mass of the qubits themselves, which is negligible, but to the surrounding classical microwave pulse hardware that delivers the bit flips. These components are typically mounted in gold-plated copper harnesses that are visibly separated in IBM's hardware layout.

Classical Bits:

- c[0]: Result of the partial measurement of the Control qubit.
- c[1]: Final measurement of the Control qubit after the delay and Hadamard.
- c[2]: Final measurement of the Gravity qubit (or in some versions, of the Test qubit).

Circuit Steps

1. Apply an X gate to the Control qubit (q0):

$$|0\rangle \rightarrow |1\rangle \quad (10)$$

2. Apply a Hadamard gate to the Control qubit (q0):

$$|1\rangle \rightarrow \frac{|0\rangle - |1\rangle}{\sqrt{2}} \quad (11)$$

3. Apply a CRY(π/4) gate from the Control qubit to an ancilla qubit to perform a partial measurement.

4. Immediately measure the ancilla to obtain a classical result c[0].
5. Conditionally flip one of the gravity qubits depending on c[0]:
   • If c[0] = 1, apply an X gate to Gravity Qubit 1 (q3).
   • If c[0] = 0, apply an X gate to Gravity Qubit 2 (q4).

This creates a spatially separated superposition in the classical control system, assuming the qubit was in an improper mixture.

6. Delay the circuit by 50 microseconds to allow time for potential objective reduction effects to manifest.
7. Apply a final Hadamard gate to the Control qubit (q0).
8. In practice, the roles of the Test and Control qubits are alternated in repeated runs to rule out qubit quality issues. Qubit coherence times (T1, T2) are pulled via the Qiskit API for each run to ensure bit quality are not being misinterpreted as collapse signals.

*Prediction*

Given a typical $T_1$ time of 300 μs and a $T_2$ time of 150 μs for IBM's Eagle processors, we can estimate the state of the system at the end of the circuit as follows:

**Step 1: Setup**

IBM Quantum computers initialize to $|0\rangle$
After applying X: $|1\rangle$
After Hadamard (H) gate

$$|\psi\rangle = \frac{|0\rangle - |1\rangle}{\sqrt{2}} \quad (12)$$

The Corresponding Density Matrix (ρ):

$$\rho = \frac{1}{2}\begin{pmatrix} 1 & -1 \\ -1 & 1 \end{pmatrix} (13)$$

Step 2: Partial Measurement (CRY Gate)

A partial measurement using a CRY gate with rotation angle π/4 reduces coherence. We approximate the post-measurement density matrix as:

$$\rho \approx \begin{pmatrix} 0.5 & -0.25 \\ -0.25 & 0.5 \end{pmatrix} \tag{14}$$

(Coherence reduced to ~0.25 from 0.5)

Step 3: Decoherence During 50 μs Delay
T1 relaxation (300 μs):
Population of |1⟩ decays exponentially:

$$P_{|1\rangle}(50) = 0.5e^{-\frac{50}{300}} \approx 0.423 \tag{15}$$

Population of |0⟩ increases correspondingly:

$$P_{|0\rangle}(50) \approx 0.577 \tag{16}$$

T2 decoherence (150 μs):
Coherence decays faster:

$$c(50) = 0.25 \times e^{-\frac{50}{150}} \approx 0.179 \tag{17}$$

Updated density matrix after 50 μs:

$$\rho \approx \begin{pmatrix} 0.577 & -0.179 \\ -0.179 & 0.423 \end{pmatrix} \tag{18}$$

Step 4: Final Hadamard Gate

Apply final Hadamard

$$H = \frac{1}{\sqrt{2}}\begin{pmatrix} 1 & 1 \\ 1 & -1 \end{pmatrix} \tag{19}$$

Calculation:

$$\rho' = H\rho H \tag{20}$$

First multiplication

$$H\rho = \frac{1}{\sqrt{2}}\begin{pmatrix} 0.577 - 0.179 & -0.179 + 0.423 \\ 0.577 + 0.179 & -0.179 - 0.423 \end{pmatrix} = \frac{1}{\sqrt{2}}\begin{pmatrix} 0.398 & 0.244 \\ 0.756 & -0.602 \end{pmatrix} \tag{21}$$

Second multiplication

$$\rho' = \frac{1}{2}\begin{pmatrix} 0.398 + 0.244 & 0.398 - 0.244 \\ 0.756 - 0.602 & 0.756 + 0.602 \end{pmatrix} = \frac{1}{2}\begin{pmatrix} 0.642 & 0.154 \\ 0.154 & 1.358 \end{pmatrix} = \begin{pmatrix} 0.321 & 0.077 \\ 0.077 & 0.679 \end{pmatrix} \tag{22}$$

Step 5: Final Measurement
The final measurement probabilities (diagonal entries) are therefore:

|0⟩: 32.1%
|1⟩: 67.9%

If objective reduction (OR) is correct, these probabilities should trend toward a 50:50 distribution, given sufficient mass and interaction time.

### *Practical Steps*

IBM Eagle computers are calibrated daily, and the individual T1 and T2 parameters can be read for every bit. We ensure that we are not observing a variation due to bit quality by accessing these values via the Qiskit API, and, more importantly, we reverse the test and control bits for each test to ensure the effect follows the gravitational difference and not the qubit quality.

### *Results*

We see a difference consistent with Penrose Objective Reduction (OR). We further see a dependence on time and mass. This dependence is not the linear one predicted by Penrose but has some non-linear structure. We are investigating this further to see if there is a quantized or oscillating relationship between collapse of the wavefunction and the gravitational interaction.

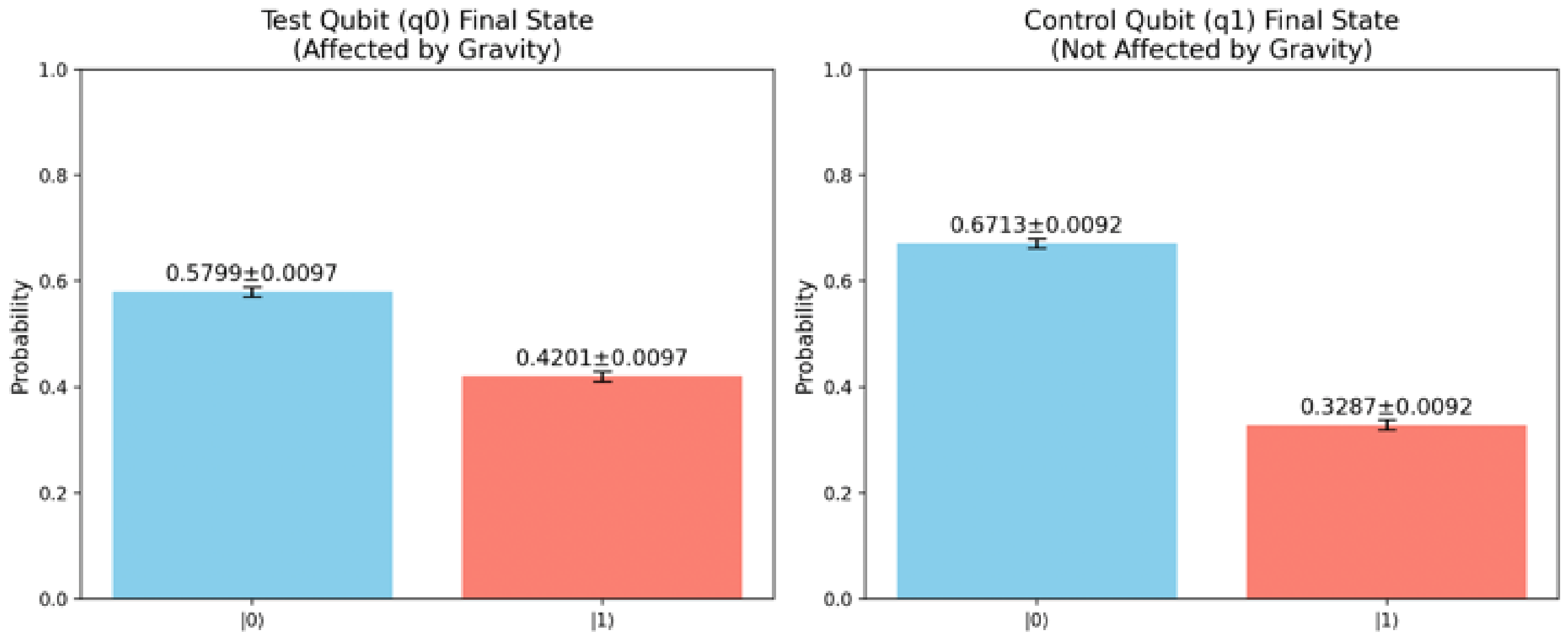

*Fig 3 shows the observed probability distributions for the Test and Control qubits across many runs. Standard quantum mechanics predicts no difference between them. Our results, however, show a statistically significant deviation:*

Test Qubit (q0, affected by gravity):
|0⟩: 57.99% ± 0.97%
|1⟩: 42.01% ± 0.97%

Control Qubit (q1, not affected by gravity):
|0⟩: 67.13% ± 0.92%
|1⟩: 32.87% ± 0.92%

*Statistically Significant Difference*:

The Test qubit shows 42.01% |1⟩ vs. the Control qubit's 32.87% |1⟩Difference: 9.14% ± 1.34% (combining error margins)

This is approximately 6.8 standard deviations, corresponding to a p-value less than 0.00001, indicating a statistically significant difference.

Closeness to 50:50 Distribution:
Test Qubit is 7.99% away from a 50:50 distribution & Control Qubit is 17.13% away.

The Test Qubit shows behavior that is closer to balanced, consistent with the prediction that gravitational superposition drives collapse toward classical probabilities.

*Conclusion*

Standard quantum mechanics does not predict any back-reaction from using the result of a partial measurement in subsequent quantum operations. The classical measurement process is expected to act as a one-way boundary—effectively isolating the quantum system from any future influence. However, our results suggest otherwise.

We observe a significant and reproducible difference when coupling a large mass to the partial measurement result via a classical bit. The result exhibits variation with time and mass. Further results will be published shortly.

*Acknowledgments*

Valis is a privately held research company. We gratefully acknowledge the support of IBM's team, who were responsive and helpful regarding using Qiskit and the Eagle system. However, the experiment used only publicly available tools and resources, and IBM has not verified it.

Our experimental code and data analysis pipeline will be released on GitHub to allow for replication and independent verification.